\documentclass{PoS}
\usepackage{notoccite}

\title{\vspace*{-5em}
	\mbox{}\hfill \mbox{\small\sc MPP-2019-247}\\
	\vspace*{6em}
	Infrared structure of $\mathcal{N}$ = 4 SYM and leading transcendentality principle in gauge theory}

\ShortTitle{Short Title for header}
\author{Taushif Ahmed\\
	Max-Planck-Institut f\"ur Physik, Werner-Heisenberg-Institut, 80805 München, Germany
	E-mail: \email{taushif@mpp.mpg.de}}
\author{Pulak Banerjee\\
	Paul Scherrer Institute, Forschungsstrasse 111, CH-5232 Villigen PSI, Switzerland\\
	E-mail: \email{pulak.banerjee@psi.ch}}
\author{\speaker{Amlan Chakraborty}\thanks{We thank the organisers of RADCOR 2019.}\\
         The Institute of Mathematical Sciences, HBNI, Taramani, Chennai 600113, India\\
        E-mail: \email{amlanchak@imsc.res.in}}

\author{Prasanna K. Dhani\\
	INFN, Sezione di Firenze, I-50019 Sesto Fiorentino, Florence, Italy\\
	E-mail: \email{prasannakumar.dhani@fi.infn.it}}
\author{V. Ravindran\\
	The Institute of Mathematical Sciences, HBNI, Taramani, Chennai 600113, India\\
	E-mail: \email{ravindra@imsc.res.in}}
\author{Satyajit Seth\\
	Institute for Particle Physics Phenomenology, Department of Physics, University of Durham, Durham, DH1 3LE, UK\\
	E-mail: \email{satyajit.seth@durham.ac.uk}}
\abstract{We present a detailed study on the infrared structure of $\mathcal{N}=4$ SYM and its connection to QCD.
	Calculation of collinear  splitting functions helps to understand the structure
	and thus one can get infrared safe cross sections.
	We also demonstrate the factorization property that soft plus virtual part of the cross section satisfies
	and through factorization, we calculate soft distribution function up to third order in perturbation theory.
	We show that the soft distribution function is process independent that includes operators
	as well as external legs. In addition to this we compare our findings against the known results in QCD
	through principle of maximum transcendentality (PMT). We extend our
	analysis further for the case of three-point
	form factors involving stress tensor and find that it violates the PMT while comparing with the corresponding quantity in the standard model, observed for the first time at the level of form factor.}

\FullConference{14th International Symposium on Radiative Corrections (RADCOR2019)\\ 
9-13 September 2019\\
		Palais des Papes, Avignon, France}

\begin{document}

\section{Introduction and theoretical framework}
\label{sec:intro}
Field theoretic  results from Quantum Chromodynamics (QCD) play an important role in understanding
the physics of strong interactions.  Inclusive and differential cross sections 
computed using perturbative QCD not only helped to discover several of elementary particles of the Standard Model (SM) but also provided a laboratory to understand the field theoretical structure of 
non-abelian gauge theories. 
For example, both theoretical and experimental results from 
		high energetic collision processes, such as the  deep-inelastic scattering and the Drell-Yan production 
		provides the complete knowledge of the internal structure as well as the dynamics of hadrons
		in terms of their constituents such as quarks and gluons. Scattering cross sections computed in high energetic collision processes such as the Drell-Yan and the  deep-inelastic scattering processes can be expressed in terms of perturbatively computed partonic cross section, convoluted with the parton distribution functions (PDFs).  The partons refer to quarks and gluons and the PDFs describe the probabilities
		of finding the partons in a bound state.These scattering cross sections at high energies can be expressed in terms of the perturbatively 
		calculable scatterings involving constituents of hadrons properly convoluted with parton distribution functions.These constituents at high energies are light quarks and gluons often called partons 
		and the corresponding PDFs describe their
		probabilities to exist in the hadron.  Such a description of hadronic cross section goes by
		the name parton model.
		Like QCD, $\mathcal{N}=4$ supersymmetric Yang-Mills (SYM) is a renormalizable gauge theory
		in four dimensional Minkowski space. In addition to having all the symmetries of QCD ,
		$\mathcal{N}=4$ SYM theory possesses 
		supersymmetry and conformal symmetry that make it
		interesting to study. Although the study of cross sections in such a theory has no 
		phenomenological implications, yet they can help us to understand 
		the factorization properties of the IR singularities, 
		the latter being useful to extract the AP kernels at each order 
		in the perturbation theory. 
		Undoubtedly, higher order computation of the FFs and the amplitudes unravel the IR structure of the $\mathcal{N}=4$ SYM
		theory in an elegant way.
		However purely real emission processes, which appear in cross sections, can also 
		give important informations about the nature of soft and collinear emissions.  In QCD, 
		the gluons in a virtual loop can become
		soft and contribute to poles in $\epsilon$ in a dimensionally regulated theory, similar 
		situation also happens when gluons in a real emission process carry a small fraction of the  momentum
		of the incoming particles. More precisely, when we perform the phase space integrations for such real
		emission processes, we encounter poles in $\epsilon$, at every order in perturbation series. 
		These soft contributions from real and virtual diagrams
		cancel order  by order when they are added together, thanks to the Kinoshita-Lee-Nauenberg (KLN) theorem \cite{Kinoshita:1962ur, Lee:1964is}.  In addition, the real emissions of gluons and quarks are sensitive to collinear singularities; 
		while the final state divergences are
		taken care by the KLN theorem, the initial state counterparts are removed by mass factorization.
		Similar scattering of  massless gluons, quarks, scalars and pseudo-scalars in ${\cal N}=4$ SYM theory can be studied within
		a supersymmetric preserving regularised scheme. 
		The cancellation of soft singularities and factorisation of collinear singularities in the scattering
		cross sections will also provide wealth of information on the IR structure of $\cal N$ = 4 gauge theory.
		One can investigate the soft plus virtual part of these finite cross sections after mass factorisation
		in terms of universal cusp and collinear anomalous dimensions. Also, the factorisation of 
		initial state collinear
		singularities provides valuable information about the AP splitting functions in $\cal N$=4 SYM theory.
		Understanding such cross sections in the light of well known results in QCD will help us to
		investigate the  resummation of soft gluon contributions to all orders in perturbation theory in a process
		independent manner.
		{\it{In this article, we focus on computing Spitting functions and try to understand the universal factorisation properties of Soft-Virtual(SV) cross-sections for certain composite operators (BPS and Stress tensor)   upto NNLO and then to predict the NNNLO  SV cross-sctions using the known three loop anamolous dimentions $\cal N$=4 SYM .}}
		
\section{Computation of splitting functions and finite cross sections}
The generic scattering process in ${\cal N} =4$ SYM theory is given by
\begin{equation}
\label{eq:process}
a(p_1) + b(p_2) \rightarrow I(q) + \sum\limits_{i=1}^{m} X(l_{i}),
\end{equation}
where $a,b \in \{\lambda,g,\phi,\chi \}$ can be a Majorana or gauge or scalar or pseudoscalar particles. 
$I$ represents a color singlet state denoted by half-BPS or Konishi or stress tensor ($T$) with invariant mass given by $Q^2=q^2$,
		where $q$ is its four momentum.   

$X$ denotes the final inclusive state comprising of 
$\{\lambda,g,\phi,\chi\}$. In the above equation, the momenta of the corresponding particles
are given inside their parenthesis  { with the invariant mass of the singlet state
	denoted by $Q^2=q^2$. Except the singlet state all other particles are massless. }

The inclusive cross section, $\hat{\sigma}^I_{ab}(\hat s,Q^2,\epsilon)$, 
for the scattering process in Eq.~(\ref{eq:process}) in $4+\epsilon$ dimensions is given by
\begin{equation}
\label{eq:sigma}
\hat{\sigma}^I_{ab}(\hat s,Q^2,\epsilon) =
\frac{1}{2 \hat{s}} \int \left[ dPS_{m+1} \right]  \overline{\sum} \left|\mathcal{M}_{ab}\right|^2,
\end{equation}
where $\hat{s} = (p_1+p_2)^2$ is the partonic center of mass energy. The phase space integration, $\int \left[dPS_{m+1}\right]$, is given by
\begin{equation}
\label{eq:phasesp}
\int \left[ dPS_{m+1} \right] = \int  \prod_{ i=1 }^{m+1} \frac{d^{n}l_i}{(2\pi)^{n}} 2\pi\delta_{+}(l_i^2-q_i^2)
(2\pi)^{n}
\delta^n
\Big( \sum_{j=1}^{m+1} l_j - p_1 - p_2 \Big)\,, 
\end{equation}
with $l_{m+1}=q$, $q_i^2 = 0$ for $i=1,\cdot \cdot \cdot m$ and $q_{m+1}^2=Q^2$.
The symbol $\overline{\sum}$ indicates sum of all the spin/polarization/generation 
and color of the final state particles $X$  and
the averaging over them for the initial state
scattering particles $a,b$. $\mathcal{M}_{ab}$ is the amplitude
for the scattering reaction depicted in Eq.~(\ref{eq:process}). 
We follow the Feynman diagrammatic approach to compute these amplitudes. 

We compute the inclusive cross section order by order in perturbation theory as
\begin{eqnarray}
\hat \sigma^I_{ab}(z,Q^2,\epsilon) = \sum_{i=0}^\infty {a}^i  \hat \sigma^{I,(i)}_{ab}(z,Q^2,\epsilon) ,
\end{eqnarray}
The UV finite virtual amplitudes involving half-BPS, T and Konishi are sensitive to IR singularities.
The massless gluons can give soft singularities and 
the massless states in virtual loops  
can become parallel to one another, giving rise to collinear singularities.
can have collinear configurations giving rise to collinear singularities.The soft singularities from the virtual diagrams cancel against the those from the real emission processes, thanks to the Kinoshita-Lee-Nauenberg (KLN) theorem~\cite{Kinoshita:1962ur, Lee:1964is}.Similarly, the final state collinear singularities cancel among themselves in these inclusive cross sections leaving only initial state collinear singularities.
{ The soft and collinear singularities  from the virtual diagrams cancel against the soft and final 
	state collinear divergences from the real emission processes, thanks to the KLN theorem}~\cite{Kinoshita:1962ur, Lee:1964is}.
Since the initial degenerate states are not summed in the scattering cross sections,
collinear divergences originating from incoming states remain as poles in $\epsilon$.  Hence,
like in QCD, the inclusive cross sections in ${\cal N}=4$ SYM theory, are singular in four dimensions. 
Following perturbative QCD~\cite{Collins:1985ue}, these singular cross sections
can be shown to factorize at the factorization scale $\mu_{F}$:
\begin{eqnarray}
\label{eq:massfact}
{\hat \Delta}^I_{ab}\left(z,Q^2,{1\over \epsilon}\right) &=&\left(\prod_{i=1}^3 \int_0^1 dx_i \right)
\delta\left(z-\prod_{i=1}^3 x_i\right)\,
~\sum_{c,d}\Gamma_{ca}\left(x_1,\mu_F^2,{1\over \epsilon}\right) 
\nonumber\\&&  
\times \Gamma_{db}\left(x_2,\mu_F^2,{1\over \epsilon}\right)
~  \Delta_{cd}^I\left(x_3,Q^2,\mu_{F}^2,\epsilon\right) ,
\end{eqnarray}
where the sum extends over the particle content $\{ \lambda, g, \phi, \chi \}$.
In the above expression $\hat{\Delta}^I_{ab}(z,Q^2,1/\epsilon) = \hat{\sigma}^{I}_{ab}(z,Q^2,\epsilon)/z$; 
the corresponding one after factorisation  
is denoted by $\Delta^I_{ab}$. 
If this is indeed the case, then we should be able to obtain $\Gamma_{ab}$ order by order
in perturbation theory from the collinear singular $\hat \Delta^I_{ab}$
by demanding $\Delta^I_{ab}$ is finite as $\epsilon \rightarrow 0$. 
The fact that the $\hat \Delta^I_{ab}$ are independent of the scale $\mu_F$ leads
the following RGE:
\begin{equation}
\label{RGkernel}
\mu_F^2 {d \over d\mu_F^2}\Gamma(x,\mu_F^2,\epsilon)={1 \over 2}  P
\left(x\right) 
\otimes \Gamma \left(x,\mu_F^2,\epsilon\right),
\end{equation}
where the function $P(x)$ is matrix valued and their elements $P_{ab}(x)$ 
are finite as $\epsilon \rightarrow 0$ and they are called splitting functions.   
{This is similar to Dokshitzer-Gribov-Lipatov-Altarelli-Parisi (DGLAP) evolution equation}~
In the  { $\overline{\rm DR}$} scheme, the solution to the RGE in terms of the splitting functions, {the latter expanded in $a$ as,}
\begin{equation}
\label{sfexpand}
{P_{{ ca}}(x) = \sum_{i=1}^{\infty} a^i P^{(i-1)}_{{ ca}}(x)},
\end{equation}
can be found to  be
\begin{equation}
\label{eq:kernel}
\Gamma_{ca}\left(x, \mu_{F}^{2},{1 \over \epsilon}\right) = \sum\limits_{k=0}^{\infty}
a^{k}\Gamma^{(k)}_{ca}\left(x, \mu_{F}^{2},{1 \over \epsilon}\right), \nonumber
\end{equation}
with
\begin{equation}
\Gamma_{ca}^{(0)} = \delta_{ca} \delta(1-x)\,,
\nonumber\\
\Gamma_{ca}^{(1)} = \frac{1}{\epsilon}  P^{(0)}_{ca}(x)\,,
\nonumber\\
\Gamma_{ca}^{(2)} = \frac{1}{\epsilon^{2}} \Bigg( \frac{1}{2} P^{(0)}_{ce} \otimes
P^{(0)}_{ea} \Bigg) + \frac{1}{\epsilon}
\Bigg( \frac{1}{2} P^{(1)}_{ca}\Bigg)\,. 
\end{equation}

Following QCD, we can relate the Mellin moments of $P_{ab}$ obtained
in ${\cal N}=4$ SYM theory with the anomalous dimensions of composite operators given by  
\begin{eqnarray}
{\cal O}^\lambda_{\mu_1\cdot \cdot \cdot \mu_j} &=&
S\left\{\overline \lambda^a_m \gamma_{\mu_1} D_{\mu_2} \cdot \cdot \cdot D_{\mu_j} \lambda^a_m\right\}\,,
\\
{\cal O}^g_{\mu_1\cdot \cdot \cdot \mu_j}&=&
S\left\{G^a_{\mu \mu_1}  D_{\mu_2} \cdot \cdot \cdot D_{\mu_{j-1}} G^{a\mu}_{\mu_j}\right\} \,,
\\
{\cal O}^{\phi}_{\mu_1\cdot \cdot \cdot \mu_j}&=&
S\left\{\phi^{a}_i D_{\mu_1} \cdot \cdot \cdot D_{\mu_j} \phi^a_i \right\}\,,
\\
{\cal O}^{\chi}_{\mu_1\cdot \cdot \cdot \mu_j}&=&
S\left\{\chi^a_i D_{\mu_1} \cdot \cdot \cdot D_{\mu_j} \chi^a_i\right\} \,.
\end{eqnarray}
The symbol $S$ indicates symmetrisation of indices $\mu_1 \cdot \cdot \cdot \mu_j$.
Note that these operators mix under renormalisation and the corresponding anomalous dimensions are
given by $\gamma_{ab,j}$.
\section {Analytical results and discussion}
We find that  both at NLO and NNLO, only the diagonal splitting functions contain ``+'' distributions.  In addition,
at NNLO level, terms proportional to $\delta(1-z)$ start contributing to diagonal splitting functions.  Hence,
{in the limit z $\rightarrow$ 1}, the diagonal splitting functions can be parametrized as
\begin{eqnarray}
\label{diagsf}
P^{(i)}_{aa}(z) = 2 A_{i+1} {1 \over (1-z)_+} + 2 B_{i+1} \delta(1-z) + R^{(i)}_{aa}(z),
\end{eqnarray}
where  $A_{i+1}$ and $B_{i+1}$ are the cusp 
\cite{Korchemsky:1987wg,Beisert:2006ez,Correa:2012nk,Ahmed:2016vgl}
and collinear \cite{Ahmed:2016vgl} anomalous dimensions respectively. $R^{(i)}_{aa}(z)$ is the regular 
function as $z\rightarrow 1$.  We find that 
\begin{eqnarray}
 A_1 = 4, A_2 = -8 \zeta_2 \,, \quad \quad and \quad \quad
 B_1 = 0, B_2 = 12 \zeta_3\,, 
\end{eqnarray} 
which are in agreement with the result obtained from the FFs of the half-BPS operator
\cite{Korchemsky:1987wg,Beisert:2006ez,Correa:2012nk, Ahmed:2016vgl}.  

Using the supersymmetric extensions of Balitskii-Fadin-Kuraev-Lipatov (BFKL) \cite{Lipatov:1976zz, Fadin:1975cb, Balitsky:1978ic}
and DGLAP  \cite{Gribov:1972ri,Lipatov:1974qm,Altarelli:1977zs,Dokshitzer:1977sg}
evolution equations,
Kotikov and Lipatov ~\cite{Kotikov:2000pm,Kotikov:2002ab,Kotikov:2004er,Kotikov:2003fb,
	Kotikov:2006ts} conjectured leading transcendentality (LT) principle 
which states that the eigenvalues of anomalous
dimension~\cite{Marboe:2016igj} matrix of twist two composite operators made out of $\lambda$, $g$ and complex $\phi$ fields 
in ${\cal N}=4$ SYM theory contain uniform transcendental terms at every order in perturbation theory. 
{Interestingly }they are related to 
{ the corresponding quantities in} QCD \cite{Moch:2004pa,Vogt:2004mw}. In~\cite{Kotikov:2004er} it has been shown that the eigenvalues of the anomalous dimension matrix are related to the universal anomalous dimension by shifts in spin-$j$ up to three-loop level. Unlike~\cite{Kotikov:2003fb}, we distinguish scalar and pseudo-scalar fields and compute their anomalous dimensions and their mixing in Mellin-$j$ space. We find two of the eigenvalues of the resulting anomalous dimension matrix coincide with the universal eigenvalues obtained in~\cite{Kotikov:2003fb} after finite shifts and the remaining two coincide with the universal ones only in the large $j$ limit (i.e. $z\rightarrow 1$). 
One can associate the transcendentality weight $n$ to terms such as $\zeta(n)$, $\epsilon^{-n}$ and
also to the weight of the harmonic polylogarithms  that appear in the perturbative calculations.
\\
\\
We now move on to study the finite cross sections $\Delta^I_{ab}$ up to NNLO level.
These cross sections
are computed in power series of the coupling constant $a$ as

	\begin{equation}
	\Delta^I_{ab} = \delta(1-z)\delta_{ab} + a~\Delta^{I,(1)}_{ab} + a^2~\Delta^{I,(2)}_{ab} + \cdot \cdot \cdot\\
	\end{equation}
	These $\Delta_{ab}^{I,(i)}$ contain both regular functions as well as 
	distributions in the scaling variable $z$.  
	The former are made up of polynomials and multiple polylogarithms of $z$ that are
	finite as $z\rightarrow 1$ and they are from hard particles.
	The distributions are from soft and collinear particles, which
	show up at every order in the perturbation theory in the form of $\delta(1-z)$ and ${\cal D}_i(z)$ 
	where
	\begin{eqnarray}
	{\cal D}_i(z) = \left( {\log^i(1-z) \over 1-z}\right)_+\,.
	\end{eqnarray}
  More precisely these distributions 
		originate from the real emission processes through
		
		\begin{equation}
		\label{matrixplus}
		(1-z)^{-1+\epsilon} = \frac{1}{\epsilon} \delta(1-z) + 
		\sum_{k=0}^{\infty} \frac{\epsilon^{k}}{k!} {\cal D}_{k}.
		\end{equation}
		These distributions constitute what is called the 
		threshold or soft plus virtual (SV) part of the cross section, denoted by 
		$\Delta_{ab}^{SV}$.
		We can  now express the total cross section as,
		\begin{eqnarray}
		\Delta_{ab}^{I,(i)} = \Delta^{I,(i),{SV}_{ab} + \Delta^{I,(i), \rm{Reg}}_{ab}},
		\end{eqnarray}
		where
		\begin{equation}
		\label{deltaSV}
		\Delta^{I,(i),SV}_{ab}= 
		\delta_{ab} \left (c^{I}_{i}\delta(1-z)+ \sum_{j=0}^{2i-1} d^I_{ij}\mathcal{D}_{j}(z) \right).
		\end{equation}
		The constants $c^{I}_{i}$ and $d^I_{ij}$ are absent when $a\not = b$.  For the diagonal ones $(a=b)$, they 
		depend on the final singlet state $I$ and are in general
		functions of rational terms and irrational $\zeta$. 
		For the diagonal ones, $\Delta_{aa}^{I,(i),\rm{SV}}$ are found identical to
		each other for $I=$\,BPS, T.  Up to NNLO level, they are found to be 
		\begin{eqnarray}
		\Delta_{aa}^{I,(0),\rm{SV}} &=&\delta(1-z)\,, \nonumber \\
		\Delta_{aa}^{I,(1),\rm{SV}} &=&8\zeta_2\delta(1-z) + 16{\cal D}_1(z)\,, \nonumber \\
		\Delta_{aa}^{I,(2),\rm{SV}} &=& 
		-{4 \over 5}\zeta_{2}^{2}\delta(1-z)
		+312\zeta_3{\cal D}_0(z) 
		- 160\zeta_2 {\cal D}_1(z)
		+ 128{\cal D}_3(z).
		\end{eqnarray}
		We observe that at every order, the above terms demonstrate uniform transcendentality which is
		{ 1 at NLO and  3 at NNLO. Note that $\delta(1-z)$ has -1 transcendental weight
			which can be understood 
			from Eq.~(\ref{matrixplus}) by noting that the term $\epsilon^{-n}$ has transcendentality
			$n$. We also notice that the highest distribution at every order determines the transcendental weight
			at that order.}
		It is interesting to note that the above coefficient functions are exactly identical to the
		LT parts of the corresponding result in the SM for the Higgs boson production through gluon fusion computed in the effective
		theory, upon proper replacement of the color factors in the following way {\it i.e.} $C_A=C_f=n_f=N$.
		On the other hand for $I=\mathcal{K}$, we find
		up to NNLO level, 
		\begin{eqnarray}
		\Delta_{aa}^{\mathcal{K},(0),\rm{SV}} &=&\delta(1-z)\,, \nonumber \\
		\Delta_{aa}^{\mathcal{K},(1),\rm{SV}} &=&\left[-28 + 8\zeta_2\right]\delta(1-z) + 16{\cal D}_1(z)\,, \nonumber \\
		\Delta_{aa}^{\mathcal{K},(2),\rm{SV}} &=& 
		\left[604 -272 \zeta_2-{4 \over 5}\zeta_{2}^{2}\right]\delta(1-z)
		+312\zeta_3{\cal D}_0(z) 
		\nonumber\\
		&&-\left[160\zeta_2 +448\right] {\cal D}_1(z)
		+ 128 {\cal D}_3(z).
		\end{eqnarray} 
		Unlike BPS and T type, for Konishi, $\Delta^{\mathcal{K},{(i)},\rm SV}_{aa}$  does not have uniform transcendentality but
		its LT terms coincide with those of BPS/T. 
		
In perturbative QCD, the fixed order predictions for the observables 
are often unreliable in certain regions of phase space due to 
the presence of large logarithms \cite{Banerjee:2018vvb}. 
For example, at the hadron colliders,
				the inclusive observables like total cross section or invariant mass distribution of finial state
				colorless state and some differential distributions contain large logarithms which can spoil
				the reliability of fixed order results.
		For example, at the partonic threshold i.e. 
				when the initial partons have just enough energy to 
				produce the final state colorless particle and soft gluons, 
				the phase space available for the gluons become severely constrained 
				giving large logarithms.  In the resummation approaches 
				these large logarithms are systematically 
				resummed to all orders in perturbation theory leading to reliable predictions. 

		The SV part of the inclusive observables in QCD is well understood to all orders in perturbation
		theory.  For example, the SV part of the inclusive cross section gets contribution from virtual part, namely 
		the form factor and the soft, collinear configurations of the real emission processes.
		In these observables, the soft singularities cancel between virtual and real emission processes, 
		while the initial  collinear ones are removed by
		mass factorisation, thus giving IR finite results. Interestingly,  
		the factorisation property 
		of these cross sections can be used to identify the process independent soft distribution function 
		which depends only the incoming states.  In addition, they satisfy certain differential equation similar
		to K+G equation of FFs.  
		The solution gives all order prediction for the soft part of the observable 
		in terms of soft anomalous dimensions $f_a$ with $a=q,g$.
		Following \cite{Ravindran:2005vv} and noting that only $\Delta^{I, \rm SV}_{aa}$ contains threshold logarithms,
		its all order structure can be expressed as
		\begin{eqnarray}
		\label{SVravi}
		\Delta^{I, {SV}_{aa}} &=&  \left(Z^{I}\left(a,\epsilon\right)\right)^2 |\hat{F}^{I}_{aa}(Q^2,\epsilon)|^2\delta(1-z)\otimes {\cal C} \exp \left(2\Phi^I_{aa}(z,Q^2,\epsilon)\right)
		\nonumber \\
		&& \otimes\Gamma^{-1}_{aa}(z,\mu_F^2,\epsilon)\otimes \Gamma^{-1}_{aa}(z,\mu_F^2,\epsilon).
		\end{eqnarray}
		In above $I$ can be any one of the three operators considered in our current work. 
		$Z^{I}(a,\epsilon)$ is the overall operator renormalization constant, which is unity
		for $I=$ half-BPS and T operators; however, for $I={\mathcal K}$, up to three loop, the pertubative coefficients of 
		$Z^{{\cal K}}$ are available ~\cite{Anselmi:1996mq, Eden:2000mv, Bianchi:2000hn,Kotikov:2004er, Eden:2004ua,Ahmed:2016vgl}. 
		$\hat{F}^{I}_{aa}(Q^2)$ is the FF contribution, {\it i.e.}, the matrix elements of
		the half-BPS or T or $\mathcal{K}$ between the on-shell state $aa$ where $a=\{\lambda,g,\phi,\chi\}$ and vacuum,
		normalised by the Born contribution, which reads as 
		\begin{eqnarray}
		\hat F^I_{aa}(Q^2) = {\langle a(p_1),a(p_2) | \tilde {\cal O}^I | 0 \rangle \over   
			\langle a(p_1),a(p_2) | \tilde {\cal O}^I | 0 \rangle^{(0)} }
		\,,\quad \quad \quad Q^2 = (p_1+p_2)^2 \,.
		\end{eqnarray}
		$\tilde{\cal O}^I$ is the Fourier transform of ${\cal O}^I$ and the superscript $0$ 
		indicates that it is the Born contribution. 
		$\Phi^I_{aa}(z,Q^2)$ is the soft distribution function resulting from the
		soft radiation and
		$\Gamma_{aa}$ are the AP kernels 
		{ that can be written in terms diagonal splitting functions as given in Eq.~(\ref{diagsf})}.

			$\Gamma_{aa}$ and they contain 
				only the distributions $\delta(1-z)$ and ${\cal D}_i(z)$ 
				from the diagonal splitting functions $P^{(i)}_{aa}(z)$. 
		The symbol $\otimes$ denotes convolution and the ${\cal C} \exp(f(z))$ is defined by
		\begin{eqnarray}
		{\cal C}e^{\displaystyle f(z) }= \delta(1-z)  + {1 \over 1!} f(z)
		+{1 \over 2!} f(z) \otimes f(z) + {1 \over 3!} f(z) \otimes f(z) \otimes f(z) 
		+ \cdot \cdot \cdot
		\end{eqnarray}
		In the above, we drop all the regular terms resulting from the convolutions and keep only distributions.
		In \cite{Ahmed:2016vgl},  the FFs are shown to satisfy the K+G equation 
		\cite{Sudakov:1954sw,Mueller:1979ih, Collins:1980ih,Sen:1981sd}
		and its solution at each order can be expressed in terms of the universal cusp ($A^I$), soft ($f^I$) and 
		collinear anomalous ($B^I$) dimensions along with 
		some operator dependent contributions \cite{Moch:2005tm,Ravindran:2005vv}. 
		$\Delta^{I, {SV}}_{aa}$ is finite in the limit $\epsilon \rightarrow 0$, thus 
		the pole structure of soft distribution function should be similar to that
		of $\hat F^I_{aa}$ and $\Gamma_{aa}$. One can show that the soft distribution function
		$\Phi^I_{aa}$ also satisfies a Sudakov type differential equation \cite{Ravindran:2005vv}
		whose solution is straightforward to obtain:
		\begin{equation}
		\label{phisoln}
		\Phi^I_{aa} = \sum_{i=1}^{\infty}a^i\bigg(\frac{q^2(1-z)^2}{\mu_{F}^2}\bigg)^{i\epsilon/2}\bigg(\frac{1}{1-z}\bigg)\bigg[\frac{2A_{i}}{i\epsilon} -f_{i} + \overline {\cal G}^I_{ia}(\epsilon) \bigg],
		\end{equation}
	
	where
	\begin{eqnarray}
	\label{softfunc}
	f_1 &=& 0\,, \quad \quad \quad f_2= -28 \zeta_3
	\quad \quad \quad f_3 =  {176 \over 3} \zeta_2 \zeta_3 + 192 \zeta_5.
	\end{eqnarray}
	We find that $\Phi^I_{aa}$ does not depend on $I$ and in addition they are identical for $a=\lambda,g,\phi$
	and $\chi$.  Hence,  $\overline{\cal G}^I_{ia} = \overline {\cal G}_i$.  From the known coefficient functions, $\Delta^{I,(i),{SV}}$, up to two loops we can determine 
	$\overline {\cal G}_i$ and they are found to be 
	\begin{eqnarray}
	\overline {\cal G}_1(\epsilon) &=& - 3\zeta_2\epsilon
	+{7 \over 3}\zeta_3\epsilon^2 
	- {3 \over 16}\zeta_2^2\epsilon^3
	+ \left[ {31 \over 20}\zeta_5
	- {7 \over 8}\zeta_2 \zeta_3 \right]\epsilon^4\, 
	\nonumber\\
	&&        + \left[{49 \over 144}\zeta_3^2
	- {57 \over 640}\zeta_2^3 \right]\epsilon^5 
	+{\cal O}(\epsilon^6),
	\nonumber\\
	\overline {\cal G}_2(\epsilon) &=&
	4\zeta_2^2\epsilon
	+43\zeta_5\epsilon^2
	+\left[{413 \over 6}\zeta_3^2
	+ {715 \over 84}\zeta_2^3 \right]\epsilon^3
	\nonumber\\
	&&        +\left[{9 \over 2}\zeta_7
	- {2527 \over 20}\zeta_2 \zeta_5
	+ {559 \over 120}\zeta_2^2 \zeta_3\right] \epsilon^4
	+{\cal O}(\epsilon^5).
	\end{eqnarray}
	The above result is found to be exactly identical to $\Phi^{q}$ and $\Phi^{g}$ that appear
	in the inclusive cross sections of the Drell-Yan and the Higgs productions respectively up to two loops,
	after setting the Casimirs of SU(N) as $C_F=n_f=C_A$ and retaining only the LT terms.  
	Our explicit computation demonstrates that the soft distribution function $\Phi$ 
	contains uniform transcendental terms and in addition it obeys leading transcendentality principle.
	In \cite{Ahmed:2014cla}, third order contribution to $\Phi^I$ for $I=q,g$ were obtained from \cite{Anastasiou:2014vaa}
	which we use here to predict the corresponding result for $\Phi$ of ${\cal N}=4$ SYM theory after suitably adjusting the 
	color factors and retaining the leading transcendental terms.  That is, we find  
	\begin{eqnarray}
	\label{gbar3}
	f_3 &=&  {176 \over 3} \zeta_2 \zeta_3 + 192 \zeta_5.
	\nonumber\\
	\overline {\cal G}_3(\epsilon) &=&
	- 4006\zeta_6
	+ {536 \over 3}\zeta_3^2
	+ {289192 \over 315}\zeta_2^3
	+{\cal O}(\epsilon).
	\end{eqnarray}
	The three-loop results for the FFs, $\hat{F}^{I}$ are already known 
	\cite{Ahmed:2016vgl}, up to the same order 
	the distribution parts of $\Gamma_{aa}$ (see Eq.~(\ref{diagsf}))  can be obtained by using
	$A_3$ \cite{Correa:2012nk,Ahmed:2016vgl} and
	$B_3$ \cite{Ahmed:2016vgl}.
	Using $f_3$ 
	and $\overline {\cal G}_3(\epsilon)$ from Eq.~(\ref{gbar3})
	we determine $\Phi^I$ up to three loops.
	Having known the form factors, soft distribution function and the AP kernels to third order,
	it is now straight forward to predict the SV part cross section at third order using Eq.~(\ref{SVravi}).  
	For $I={\cal K}$, we find  
	\begin{eqnarray}
	\Delta_{\phi\phi}^{\mathcal{K}, (3),{SV}} &=&
	\left[
	- {8012 \over 3}\zeta_6
	+ {13216 \over 3}\zeta_3^2
	+ 480 \zeta_5
	- {992 \over 5} \zeta_2^2
	- 432 \zeta_3
	+ 6512\zeta_2
	- 11552
	\right]\delta(1-z)
	\nonumber\\
	&&    
	+ \left[
	11904\zeta_5
	- {23200 \over 3}\zeta_2 \zeta_3
	- 8736\zeta_3
	\right] {\cal D}_0
	+   \left[
	- {9856 \over 5}\zeta_2^2
	+ 3712\zeta_2+9664
	\right] {\cal D}_1
	\nonumber\\ && 
	+   
	11584\zeta_3
	{\cal D}_2
	+   \left[
	- 3584\zeta_2
	- 3584
	\right] {\cal D}_3
	+   
	512
	{\cal D}_5.
	\end{eqnarray}
	and for the $I=$ half-BPS and T, we find 
	\begin{eqnarray}
	\label{thirdBPST}
	\Delta^{I,(3),\rm{SV}}_{aa} = \Delta^{\mathcal{K},(3),\rm{SV}}_{\phi\phi} \Big |_{\rm{LT}}\,,\quad \quad \quad 
	\end{eqnarray}
	{where for $I=$ half-BPS, $a=\phi$ and for $I=$ T, $a=\{\lambda,g,\phi,\chi\}$}.
	In addition we find that for $I=$ half-BPS, our third order prediction, Eq.~(\ref{thirdBPST}), agrees with the result 
	\cite{Li:2014afw}
	obtained by explicit computation. 
	\section{Principle of maximal transcendentality and its violation}
	\label{sec:PMT}
	The principle of maximal transcendentality (PMT) ~\cite{Kotikov:2001sc,Kotikov:2004er,Kotikov:2006ts} establishes a bridge between the results in QCD and those of ${\cal N}=4$ SYM which are comparatively simpler. It states that for certain quantities, the results in ${\cal N}=4$ SYM can be obtained from that in QCD by converting the fermions from fundamental to adjoint representation $\{C_F \rightarrow C_A, 2 n_f T_F \rightarrow C_A\}$ and then retaining only the leading transcendental (LT) weight terms. The complete domain of applicability of this principle is still not clear and under active investigation. We have compared three-point Form Factor(FF) of stress tensor operator for various external states with available QCD results for stress tensor~\cite{Ahmed:2014gla}. The LT part of the QCD results under transformation  $C_F \rightarrow C_A, 2 n_f T_F \rightarrow C_A$ do not match with ${\cal N}=4$ SYM~\cite{Ahmed:2019nkj}. This indicates a clear violation of the PMT. This non-matching may be attributed to the incomplete factorisation of the leading order amplitude from the one loop form factor of stress tensor in QCD.

	In the context of two-point two-loop form factors (FFs) with two identical operators insertion of supersymmetry protected half-BPS primary and unprotected Konishi, the PMT which dictates the presence of identical highest weight terms in the scalar FFs of half-BPS and quark/gluon FFs in QCD is found to be violated as well~\cite{Ahmed:2019yjt}.  The LT weight terms of double half-BPS FFs do not match with that of di-Higgs production
through gluon fusion or bottom quark annihilation, observed for the first time.

	\section{Conclusion}
	\label{conclusion}
	In this article, we have studied the perturbative structure of ${\cal N}=4$ SYM gauge theory  
	in the infrared sector.  We have achieve this by computing various 
	inclusive scattering cross sections of on-shell particles belonging to this theory. We have computed splitting functions up to NNLO in perturbation theory and thus found infrared safe Soft-virtual (SV) cross-section for various composite operators. We have studied the universal factorisation properties of the SV cross-section and thus predicted the N$^3$LO SV  cross-sections using the known anomalous dimensions.

\bibliography{n4}
\bibliographystyle{JHEP}

\end{document}